\documentclass[twocolumn,showpacs,preprintnumbers,amsmath,amssymb]{revtex4}
\usepackage{graphicx,bm}
\usepackage{color}

\begin{document}

\title{Numerical simulation of time delays in light induced ionization}

\author{Jing Su}
\affiliation{JILA and Department of Physics, University of Colorado, Boulder, CO 80309-0440, USA}
\author{Hongcheng Ni}
\affiliation{JILA and Department of Physics, University of Colorado, Boulder, CO 80309-0440, USA}
\author{Andreas Becker}
\affiliation{JILA and Department of Physics, University of Colorado, Boulder, CO 80309-0440, USA}
\author{Agnieszka Jaro\'n-Becker}
\affiliation{JILA and Department of Physics, University of Colorado, Boulder, CO 80309-0440, USA}

\begin{abstract}
We apply a fundamental definition of time delay, as the difference between the time a particle spends
within a finite region of a potential and the time a free particle spends in the same region, to
determine results for photoionization of an electron by an
extreme ultraviolet (XUV) laser field using numerical simulations on a grid. Our numerical results
are in good agreement with those of the Wigner-Smith time delay, obtained as the derivative of the
phase shift of the scattering wave packet with respect to its energy, for the short-range Yukawa potential.
In case of the Coulomb potential we obtain time delays for any finite region, while - as expected -
the results do not converge as the size of the region increases towards infinity. The impact of an ultrashort
near-infrared probe pulse on the time delay introduced here is analyzed for both the Yukawa as well as the Coulomb potential and is found to be small for intensities below $10^{13}$ W/cm$^2$.
\end{abstract}

\pacs{33.80.Rv, 33.80.Wz}

\date{\today}

\maketitle

\section{Introduction}
The development of attosecond extreme ultraviolet (XUV) laser technology in recent years
has offered the opportunity to observe and control the dynamics of electrons and the coupling
to nuclear dynamics in atoms and molecules on their natural time scale.
In particular, the capability to lock XUV pulses to a near-infrared (near-IR)
pulse has initiated the development of techniques in which the dynamics is triggered
by the attosecond pulse and observed as a function of the delay between
the XUV and the near-IR pulses. Experimental observations include, among others, the time resolution of the Auger decay
\cite{drescher02}, the dynamics of electrons in valence shell \cite{goulielmakis10} and excited states \cite{uiberacker07,gagnon07}, shake-up
processes \cite{uiberacker07}, and delays in the photoemission of electrons from different bands
in a solid \cite{cavalieri07} or different sub-shells in an atom \cite{schultze10,klunder11}.

In particular, observations of substantial time delays during photoionization of atoms have
generated significant theoretical interest (e.g., \cite{schultze10,klunder11,kheifets10,nagele11,ivanov11,nagele12,zhang11,zhang10,kheifets11,ivanov11pra,ivanov12,sukiasyan12,dahlstrom12,spiewanowski12,guenot12,pazourek12,moore11}). These measurements are often analyzed in terms of the so-called
Wigner-Smith (WS) time delay (e.g., \cite{schultze10,klunder11,kheifets10,nagele11,ivanov11,nagele12,zhang11,dahlstrom12}). 
The WS time delay accounts for the delay in the
propagation of a particle in a potential as compared to that of a corresponding free particle 
towards infinity in space
%over a distance $R$
in an atomic or molecular scattering scenario \cite{wigner55,smith60}.
%In the latter context i
It has been pointed out \cite{smith60}, that this definition leads to
a well-defined time delay 
%as $R \rightarrow \infty$ 
as long as the potential vanishes quickly enough
at large distances. In contrast, 
for long-range potentials, such as the Coulomb potential, the WS time delay is
an intrinsically ill-defined concept. 
In view of this deficiency of the WS time delay concept, sometimes short- and long-range parts 
of a potential are considered separately (e.g., \cite{nagele11,ivanov11,dahlstrom12}). For a given problem it may however be unclear where such a separation
is justified. Furthermore, the WS time delay is often calculated via the derivative of the
phase shift of the wavefunction with respect to the energy of the particle
(e.g., \cite{kheifets10,nagele11,ivanov11,zhang11,dahlstrom12,wigner55,smith60}).This time-independent approach does not enable an analysis 
of the delay as a function of time during the interaction.
%appears to be not always
%applicable to the photoionization process.

We therefore seek for an alternative time-dependent theoretical approach to calculate time delays 
in photoionization, which addresses some of the concerns regarding the WS time delay and its determination 
via the phase derivative outlined in the previous paragraph. We further attempt to apply such an approach in time-dependent 
numerical grid simulations which are known to be a powerful tool in calculating and analyzing processes 
on an ultrashort time scale. The present theoretical analysis of a time delay is intended to be general and not focused, in particular, on the recent streaking experiments. Once formulated, tested and established this may turn out in future as a useful step towards understanding the physics of time delays in streaking experiments and other precise measurements of ultrashort time scales.

Our proposal is based on the quantum mechanical expression for the time a particle spends
inside a certain region $R$ of a potential. By comparing this time to the corresponding time for a free particle,
a time delay is given, which is well-defined for any finite region. This approach is also known to be
the basis for the WS time delay itself, that is nothing else than the limit, if it exists, as the region $R$ grows to infinity \cite{wigner55,smith60}.
To the best of our knowledge, this fundamental definition of a time delay has not been applied in the 
analysis of time-dependent processes initiated or driven by ultrashort laser pulses. 
It however offers a few interesting features: First, as mentioned above, for any
finite region $R$ the time delay
is well-defined for any physical relevant potential and independent whether or not the limit for an extension
of the region towards infinity exists. This enables a theoretical analysis in particular for long range potentials
without any restriction of the potential.
Second, in the limit to infinity, if well-defined, the time delay should converge to the WS delay. Third, the time delay can be determined as a function of time after the emission of the
photoelectron, in case of a streaking experiment even during the interaction with the probe pulse. This expands the 
options for a theoretical analysis of ultrashort time-dependent processes. Fourth, there is no
a-priori separation of short- and long-range parts in the potential necessary and the influence of both contributions
can be studied.

In this paper we present and discuss the application of the above-mentioned time delay concept to the photoionization
process. We further show how the concept can be utilized in time-dependent numerical simulations on a grid
using the back-propagation technique. While the theoretical approach is developed and formulated in 3D, we
restrict ourselves to an implementation in 1D calculations. This is done in the present proof-of-principle study in order
to carefully investigate the convergence of the results with respect to the grid parameters, which appears to be appropriate
in view of the required resolution of time delays on the order of a few attoseconds. An application of the approach
to more dimensions is straightforward. We also show that in our application of the concept the numerical results
indeed agree with the WS time delay, in case the appropriate limit exists. We may emphasize that any
time delay determined in the present context is well-defined, even in case of the Coulomb potential, since we
consider delays over finite ranges in space only. Besides the introduction of this complementary concept and the
demonstration of its application to photoionization, we consider one
%Having this in mind we note that an important
aspect of the recent observations of time delays
using attosecond XUV pump and near-IR probe pulses, namely
the impact of the probe pulse on the time delay introduced here. We may note that this time delay does not necessarily correspond to or fully include the time delay observed in recent streaking experiments, since our method calculates (or measures) the time delay directly in the time domain while, in contrast, in the streaking technique a time delay is determined indirectly via a momentum (or energy) measurement.
%result of the measurement
Since the probe pulse is usually ultrashort, the effect
occurs over a finite time and, hence, during the propagation of the electron over a finite distance $R$
in the potential only.
Thus the present concept appears to be well suited and can therefore be
%Since the above mentioned time delay is a well-defined quantity over any
used to analyze the impact of the probing pulse for
short- as well as long-range potentials.
%Furthermore, it is possible to study the effect using
%time-dependent numerical simulations on a grid, which intrinsically need to incorporate finite times and distances
%as well.

The paper is organized as follows.
%In this article we report about such an analysis for photoionization of an electron in the short-range
%Yukawa potential as well as in the long-range Coulomb potential. To this end, 
We first provide the
basic definitions for the calculation of a time delay with and without a strong probing field and
discuss the application in numerical simulations. We then show that the numerical results
for single XUV photoionization are well-defined over finite distances but clearly do not converge
as $R \rightarrow \infty$ in the Coulomb case, in agreement with the discussion given in the
early work by Smith \cite{smith60}. On the other hand, our numerical results are in good agreement with
those for the original WS time delay in case of a short-range potential.
Finally, we investigate the impact of an ultrashort near-IR probing pulse on the results for the time delay. Our
results indicate that the effect is small as long as the intensity of the probe pulse does not exceed
$10^{13}$ W/cm$^2$.

\section{Theoretical method}

In this section we introduce the theoretical method, which we use
to obtain time delays associated with the ionization of a target system in numerical simulations.
To this end, we will first provide a set of basic definitions used in the method before we discuss
its use in numerical simulations.

\subsection{Basic definitions}

For a particle in a given state $\Psi({\bm r},t)$ the time spent inside a region $R$ with a potential $V({\bm r})$ can be expressed as
(Hartree atomic units, $e = m = \hbar = 1$ are used throughout the paper) \cite{newton_book}:
\begin{equation}
t_{\Psi,R} = \int_{-\infty}^{\infty}dt \int_R d{\bm r} |\Psi({\bm r},t)|^2\; .
\label{time_region}
\end{equation}
While $t_{\Psi,R}$ is, in general, finite for finite regions and any $\Psi({\bm r},t)$, it is useful to compare $t_{\Psi,R}$ to the time spent
by the free particle in $R$ (or another reference time):
\begin{equation}
t_{\Psi^{(0)},R} = \int_{-\infty}^{\infty}dt \int_R d{\bm r} |\Psi^{(0)}({\bm r},t)|^2.
\end{equation}
Here, $\Psi^{(0)}({\bm r},t)$ is the free particle state corresponding to $\Psi({\bm r},t)$.
The difference between $t_{\Psi,R}$ and $t_{\Psi^{(0)},R}$ defines the time delay associated with $\Psi({\bm r},t)$, the region $R$ and the potential $V({\bm r})$:
\begin{equation}
\Delta t_{\Psi,R} = t_{\Psi,R} - t_{\Psi^{(0)},R}\; .
\label{num_delay}
\end{equation}
The quantity $\Delta t_{\Psi,R}$ is known to have a finite limit as the radius of $R$ grows to infinity if the interaction vanishes quickly enough \cite{smith60}.
Thus, $\Delta t_{\Psi,R\rightarrow\infty}$ (and the associated quantum mechanical operator) is well-defined for short-range potentials $V({\bm r})$ only.
Provided that the limit exists, $\Delta t_{\Psi,R\rightarrow\infty}$ can be also expressed as the energy derivative of the phase shift $\varphi$ induced by the potential $V({\bm r})$:
\begin{equation}
\Delta t_{\Psi,R\rightarrow\infty} = \Delta t_{\text{WS}}= \frac{d\varphi}{dE},
\label{WS_delay}
\end{equation}
which is commonly known as the Wigner-Smith (WS) time delay.

The definition given above provides a useful concept to calculate time delays in time-dependent 
processes, in particular on an ultrashort time scale. While it is known as the basis for a derivation of the WS time 
delay in scattering scenarios, to the best of our knowledge it has not been applied for the theoretical analysis of 
processes initiated or driven by ultrashort intense laser pulses. As an application, we intend to obtain time delays in the form of the time difference, given in Eq.\ (\ref{num_delay}), for a photoionization process.
Physically, we are interested in the time that an - initially bound - electron needs to leave a certain region (centered about the location of the residual target) following ionization
due to the interaction with an external light field. To this end, we note that the expressions above can be readily applied to a particle in a superposition of states and
therefore consider the ionizing part (i.e.\ the continuum parts) of the wavefunction
$\Psi_i^{(\text{ion})}({\bm r},t)$ in our adoption of Eq.\ (\ref{time_region}):
\begin{equation}
t_{\Psi_{i},R} = \frac{1}{P_\text{ion}}\int_{-\infty}^{\infty}dt \int_R d{\bm r} |\Psi_i^{(\text{ion})}({\bm r},t)|^2,
\label{time_ion}
\end{equation}
where $P_\text{ion}=\int_{-\infty}^{\infty}d{\bm r}|\Psi_i^{(\text{ion})}({\bm r},t\rightarrow\infty)|^2$ is the ionization probability.
We renormalize the wavefunction via division by $P_\text{ion}$ in Eq. (\ref{time_ion}) in order to be able to compare times and time delays arising for the ionization from different initial bound states $\Psi_i({\bm r},t=0)$.
We can then define the time delay associated with the ionization from a specific initial bound state analogous to Eq.\ (\ref{num_delay}) as:
\begin{equation}
\Delta t_{\Psi_i,R} = t_{\Psi_i,R} - t_{\Psi_{i}^{(0)},R},
\label{ion_delay}
\end{equation}
where $\Psi_{i}^{(0)}({\bm r},t)$ is the free particle state
corresponding to the ionizing part of the wavefunction after transition from the initial state $\Psi_i({\bm r},t=0)$.
According to this definition we expect negative values for the time delays, since a free wave packet should
spend more time in a given region $R$ than the corresponding wave packet that has the same asymptotic energy propagating in an attractive potential.
We expect that $\Delta t_{\Psi_i,R}$ has a well-defined finite limit (for $R \rightarrow \infty$), i.e.\ the Wigner-Smith time delay,
for short-range potentials, but not necessarily for long-range potentials
such as the Coulomb interaction. In view of the intrinsic negative time delays for finite regions, we expect that the limit value
is negative as well.
We also consider the difference in the time delays for the ionizations from
two different initial states $\Psi_i({\bm r},t=0)$ and $\Psi_j({\bm r},t=0)$ as:
\begin{equation}
\Delta T (\Psi_i,\Psi_j;R) = \Delta t_{\Psi_i,R} - \Delta t_{\Psi_j,R}.
\end{equation}

\subsection{Numerical simulations of time delays}

In order to use the above definitions in a numerical simulation of a photoionization process we need to identify the ionizing part of the
wavefunction $\Psi_i^{(\text{ion})}({\bm r},t)$ as well as the corresponding free particle state $\Psi_i^{(0)}({\bm r},t)$.
Since it is not straightforward to obtain e.g.\ the time of ionization (e.g.\ \cite{kheifets10})
and the form of the wave packet after the transition into the continuum, it appears to be difficult to make use of Eq.\ (\ref{ion_delay})
in a numerical simulation directly. We circumvent this obstacle by using the back-propagation technique.

We first solve the time-dependent Schr\"odinger equation (TDSE) of the system, initially in the state $\Psi_i({\bm r},t=0)$, under the interaction
with the external light field on a space-time grid:
\begin{equation}
i\frac{\partial}{\partial t} \Psi({\bm r},t)
=
\left(\frac{{\bm p}^2}{2} + V({\bm r}) + V_{\text{light}}(t)\;
\right)
\Psi({\bm r},t),
\label{schroedinger}
\end{equation}
where ${\bm p}$ is the momentum operator and $V_{\text{light}}(t)$ represents the interaction with the ionizing light field.
After the end of the interaction with the light field we separate the ionizing part of the wavefunction from the remaining bound parts, either via
projection onto analytically or numerically known states or via spatial separation of the ionizing part at
large distances on the grid. After removal of the bound parts we propagate the remaining ionizing part of the wavefunction backwards
in time without taking account of the interaction with the light field using two different Hamiltonians, once including the potential $V({\bm r})$:
\begin{equation}
i\frac{\partial}{\partial t} \Psi_i^{(\text{ion})}({\bm r},t)
=
\left(\frac{{\bm p}^2}{2} + V({\bm r})
\right)
\Psi_i^{(\text{ion})}({\bm r},t),
\end{equation}
and once as a free particle:
\begin{equation}
i\frac{\partial}{\partial t} \Psi_i^{(0)}({\bm r},t)
=
\frac{{\bm p}^2}{2}
\Psi_i^{(0)}({\bm r},t).
\end{equation}
In order to calculate the time delay $\Delta t_{\Psi_i,R}$ for a given region $R$, the wavepacket has to be located
outside of $R$ at the start of the back-propagation and the propagation needs to be terminated as the wavepacket reaches the center
of $R$, i.e.\ the location of the residual target ion. The latter point will be further discussed in the application of the method
below.

\section{Application to single photoionization by an XUV pulse}

\subsection{Model systems}

The theoretical method outlined above is, in general, applicable to ionization of an atom or molecule in any light field.
Here we present results for the application to photoionization of an electron initially bound in
two different model potentials.
First, we used a short-range Yukawa potential in 1D:
\begin{equation}
V_{\text{Y}}(x)=-\frac{Z}{\sqrt{x^2+a}}e^{-\frac{|x|}{b}},
\label{Yukawa}
\end{equation}
where $Z$ is the effective nuclear charge, $a$ is the soft-core parameter, and $b$ is a parameter that determines the effective range of this 1D-potential.
For our simulations we chose $Z=3.0$, $a=2.0$, and $b=30.0$, which relate to energies of $-1.6742$ a.u.\ and $-1.0124$ a.u.\ of the ground and first
excited states.
As a long-range interaction we made use of the Coulomb potential in 1D:
\begin{equation}
V_{\text{C}}(x)=-\frac{Z}{\sqrt{x^2+a}}.
\label{Coulomb}
\end{equation}
For $Z=3.0$ and $a=2.0$, the energies of the lowest two states are $-1.7117$ a.u.\ and $-1.0807$ a.u.,
which are close to the energies of the Yukawa potential.

For the interaction with the XUV light pulse we used length gauge, i.e.
\begin{equation}
V_{\text{light}}(t) = E_{\text{XUV}}(t) x
\end{equation}
where $E_{\text{XUV}}$ represents a linearly polarized pulse with a sin$^2$ envelope, i.e.\
\begin{equation}
E_{\text{XUV}}(t)=E_0 \sin^2(\pi t/\tau) \sin(\omega t+\phi),
\label{XUV_pulse}
\end{equation}
where $E_0$ is the peak amplitude, $\tau$ is the pulse duration, $\omega$ is the central frequency, and $\phi$ is the carrier-envelope phase (CEP).

To solve the corresponding TDSE, we used the common Crank-Nicolson method in a grid representation.
In general, we used a spatial step of
$\delta x =0.02$ and a time step of
$\delta t=0.002$ in our simulations.
The grid extended from $-4000$ a.u.\ to $4000$ a.u.\ for the numerical simulations of the model systems
interacting with an XUV pulse
to hold the full wavefunction on the grid. The initial ground and first excited states were obtained by
imaginary time propagation method.
We continued the propagation of the wavefunction after the interaction with the XUV pulse until the ionizing
parts of the
wavepacket reached a distance beyond $|x|\geqslant 500$ and hence were well separated from the remaining bound parts.
This allowed us to remove the latter parts from the grid and remain the ionizing parts of the wavefunction only.
We then propagated the ionizing parts at negative and positive $x$ backwards in time independently, either
under the influence of the potential, $V_Y$ or $V_C$, or as
a free particle. We determined the corresponding times $t_{\Psi_i,R}$ and $t_{\Psi_{i}^{(0)},R}$ for both parts
of the ionizing wavefunction and added the two contributions.
In the 1D calculations we defined
the region as $R = [\pm x_\text{inner}, \pm x_\text{outer}]$, where
$x_\text{inner}$ and $x_\text{outer} \le 500$ are the inner and outer
boundaries, respectively, and the $\pm$-signs apply to back-propagation of
the two parts of the ionizing wavepacket along the positive/negative $x$-axis, respectively.
We absorbed the wavefunction beyond the inner boundary $x_\text{inner}$ using the exterior complex scaling method \cite{ho83,mccurdy04}.

\subsection{Boundaries and grid parameters}

\begin{table}[t]
\caption{
Results of numerical calculations for the
times $t_{\Psi_g,R}$, $t_{\Psi_g^{(0)},R}$ and the time delay $\Delta t_{\Psi_g,R}$
for different spatial steps $\delta x$ and a fixed time step of $\delta t=0.002$.
Results are obtained for ionization from the ground state of the 1D Yukawa potential and
$R=[0,\pm 460]$.
The parameters of the XUV pulse were:
peak intensity $I=1\times10^{15}$ W/cm$^2$,
frequency $\omega=100$ eV,
pulse duration $\tau=400$ as, and
carrier-envelope phase $\phi=0$. } % title of Table
\centering % used for centering table
\begin{tabular}{cccc} % centered columns (4 columns)
\hline %inserts double horizontal lines
spatial step&$t_{\Psi_g,R}$&$t_{\Psi_{g}^{(0)},R}$&$\Delta t_{\Psi_g,R}$\\ [0.5ex] % inserts table
%heading
\hline % inserts single horizontal line
$0.5$&$275.3389$&$274.4569$&$0.8820$\\ [0.5ex] % inserts table
$0.2$&$240.6028$&$241.5927$&$-0.9899$\\ [0.5ex] % inserts table
$0.1$&$236.8736$&$237.9043$&$-1.0307$\\ [0.5ex] % inserts table
$0.05$&$235.9751$&$237.0109$&$-1.0358$\\ [0.5ex] % inserts table
$0.02$&$235.7258$&$236.7628$&$-1.0370$\\ [0.5ex] % inserts table
$0.01$&$235.6903$&$236.7274$&$-1.0371$\\ [0.5ex] % inserts table
\hline %inserts single line
\end{tabular}
\label{convergence_dx_change}
\end{table}

\begin{table}[t]
\caption{
Results of numerical calculations for the
times $t_{\Psi_g,R}$ and $t_{\Psi_g^{(0)},R}$ and the time delay $\Delta t_{\Psi_g,R}$
for different time steps $\delta t$ and a fixed spatial step of $\delta x=0.02$.
All the other parameters were the same as in Table I.
} % title of Table
\centering % used for centering table
\begin{tabular}{cccc} % centered columns (4 columns)
\hline %inserts double horizontal lines
time step&$t_{\Psi_g,R}$&$t_{\Psi_{g}^{(0)},R}$&$\Delta t_{\Psi_g,R}$\\ [0.5ex] % inserts table
%heading
\hline % inserts single horizontal line
$0.1$&$237.0072$&$238.0431$&$-1.0359$\\ [0.5ex] % inserts table
$0.05$&$236.0391$&$237.0757$&$-1.0366$\\ [0.5ex] % inserts table
$0.02$&$235.7752$&$236.8121$&$-1.0369$\\ [0.5ex] % inserts table
$0.01$&$235.7378$&$236.7747$&$-1.0369$\\ [0.5ex] % inserts table
$0.005$&$235.7284$&$236.7654$&$-1.0370$\\ [0.5ex] % inserts table
$0.002$&$235.7258$&$236.7628$&$-1.0370$\\ [0.5ex] % inserts table
$0.001$&$235.7254$&$236.7624$&$-1.0370$\\ [0.5ex] % inserts table
\hline %inserts single line
\end{tabular}
\label{convergence_dt_change}
\end{table}

Based on the results of recent observations \cite{schultze10,klunder11} and
calculations \cite{kheifets10,nagele11,ivanov11,moore11,nagele12}, we expect that the time delay $\Delta t_{\Psi_i,R}$
as well as the difference in the time delays for the ionization from different
initial states $\Delta T(\Psi_i,\Psi_j,R)$ are of the orders of a few tens of attoseconds. 
Resolution of such small times
requires 
%sufficiently small 
an analysis of the time and spatial steps in the numerical simulations in order to establish 
appropriate limits for grid parameters towards a convergence of the results in the present studies.
In Tables \ref{convergence_dx_change} and \ref{convergence_dt_change} we present a set of numerical results
obtained for different $\delta x$ and $\delta t$ in the case of the Yukawa potential.
We see that a convergence of the time delay $\Delta t_{\Psi_i,R}$ within less than $0.001$ a.u. (i.e.\ $<0.025$ as)
is reached for a time step of $\delta t=0.002$ and a spatial step of $\delta x=0.02$. Similar conclusions
hold for our studies with the Coulomb potential as well.

\begin{figure}[t]
  \begin{center}
  \includegraphics[scale=0.45]{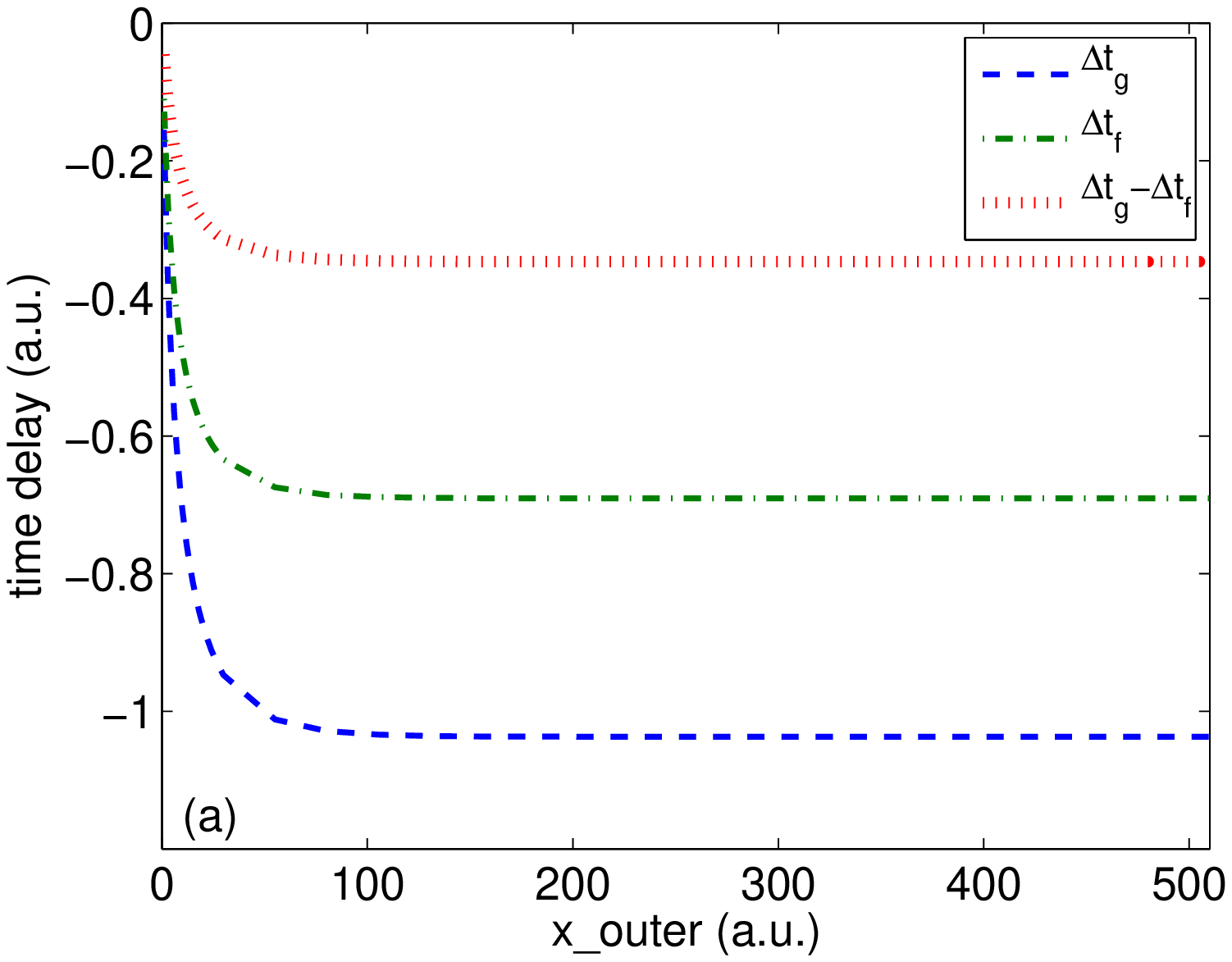}
  \includegraphics[scale=0.45]{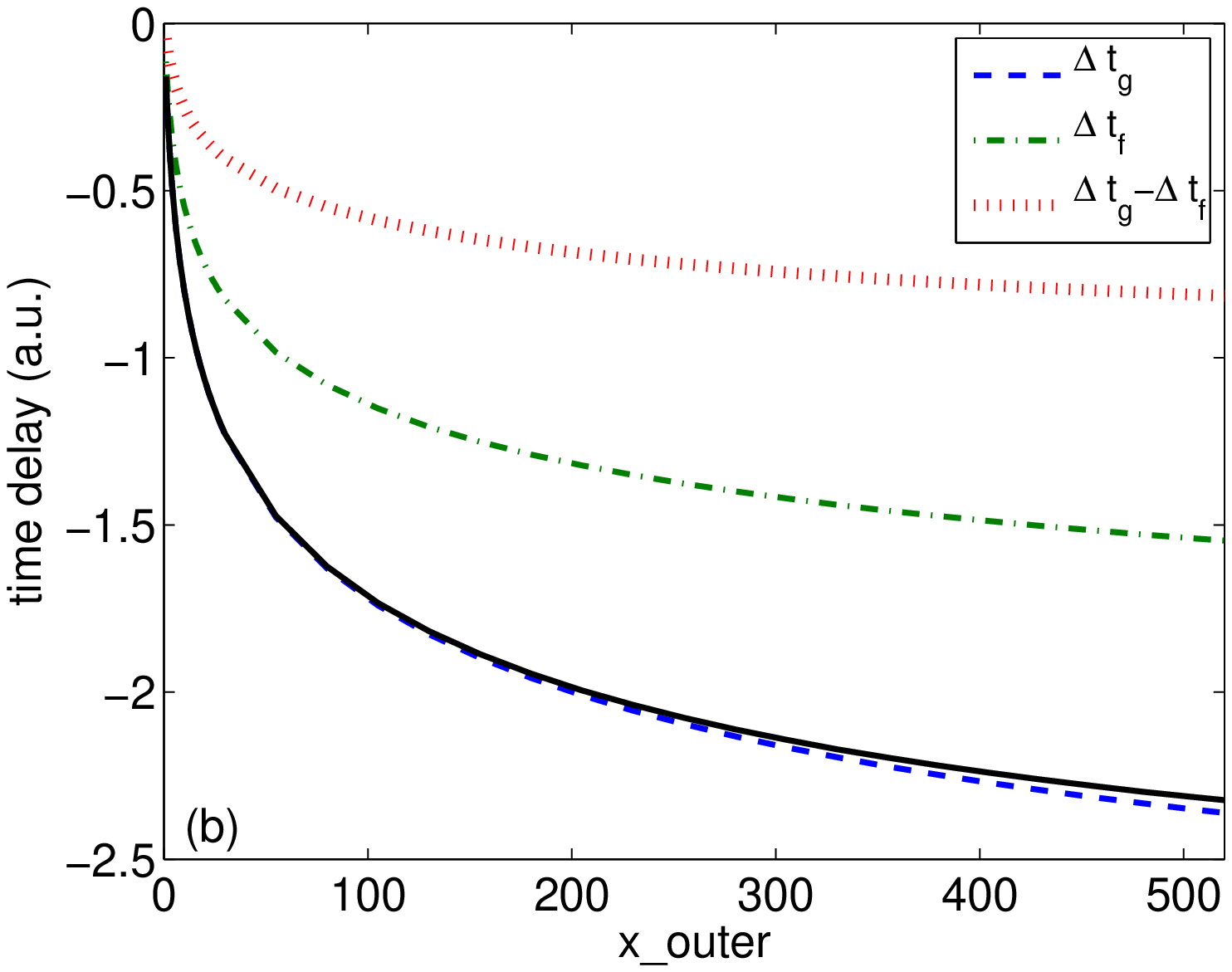}
  \end{center}
  \caption{\small
(Color online) Time delays $\Delta t_{\Psi_i,R}$ and time difference $\Delta T(\Psi_i,\Psi_j,R)$
as a function of the outer integration boundary $x_\text{outer}$ for two potentials:
(a) short-range Yukawa potential and (b) long-range Coulomb potential.
Time delays obtained for the ground and first-excited states are represented by blue dashed lines and green dash-dotted lines,
respectively; while the red dotted lines show the results for the time difference between the delays. In (b) the black solid and blue dashed lines correspond to two different forward propagation distances: $\langle x_\text{forward}\rangle=2000$ and $3000$, respectively, for the ionization from the ground state.
In all calculations we have used an XUV pulse with peak intensity $I=1\times10^{15}$ W/cm$^2$,
central frequency $\omega=100$ eV, pulse duration $\tau=400$ as, and carrier-envelope phase $\phi=0$
for the ionization.
}
\label{delay_outer_boundary}
\end{figure}

As mentioned above, the time delay $\Delta t_{\Psi_i,R}$ depends on the size of the region $R$, and should be negative and converge
to a finite limit for short-range potentials only. To test these expectations, we performed a set of simulations for
the time delays for photoionization from the
ground and excited states of both potentials as a function of the outer boundary
$x_\text{outer}$ by fixing $x_\text{inner}=0$, i.e.\ for $R=[0,x_\text{outer}]$.
As expected, the values for the time delays are negative and
decrease for an increase of the region $R$ for each of the results presented in Fig.\ \ref{delay_outer_boundary}.
For the Yukawa potential (panel a) convergence
is found for outer boundaries $x_\text{outer} > 150$.
Consequently, for large values of the outer boundary we obtain a well-defined value for the time difference $\Delta T$ of the time
delays for ionization from the ground and the excited states.

In contrast, our results do not show a convergence for the time delays as a function of the outer boundary in case
of the long-range Coulomb potential (see Fig. \ref{delay_outer_boundary}(b)). This reflects the well-known logarithmic
divergence of the time delay for this kind of potential and, hence, for ionization from any bound state within the
potential. Of course, in these cases a WS time delay as the derivative of the phase shift
(c.f. Eq. (\ref {WS_delay})) cannot be defined as well, since its derivation requires a finite limit of
$\Delta t_{\Psi_i,R\rightarrow\infty}$. It is interesting to point out that the results in
Fig. \ref{delay_outer_boundary}(b) further show that the logarithmic divergence is, in general, still present for
the difference between a pair of time delays obtained for the ionization from two different initial states.
Thus, such a time delay difference does not simply depend on the
short-range character of the potential but contains information
about the long-range part of the potential and is therefore not well-defined as well.
The present results agree well with the conclusions of early works on time delays \cite{smith60}.
%However, our results for the time delays for any finite region are finite and therefore allow us to
We may however reemphasize that any time delay obtained for a finite region via the present method is finite and therefore well-defined, even in the case of the long-range Coulomb potential. As pointed out above this allows us to
study certain aspects with respect to the parameters of the XUV pulse and the effects of an IR streaking in the
Coulomb case.

It is necessary to point out that for the Coulomb potential the time delay also depends on the distance that the ionizing wave packet
is propagated in forward direction. This is due to the long-range character of the Coulomb potential since the central momentum of the ionizing wave packet decreases with an increase of the forward propagation distance. Thus, the velocity of the free particle during the back propagation decreases as well. In Fig.\ \ref{delay_outer_boundary}(b) we show this effect by presenting results for the time delay from the ground state for two forward propagation distances: $\langle x_\text{forward}\rangle=2000$ (black solid line) and $3000$ (blue dashed line). As expected, the time delays for $\langle x_\text{forward}\rangle=3000$ are slightly smaller than those for $\langle x_\text{forward}\rangle=2000$. This shows the need to use rather large grids for the numerical simulations in case of a Coulomb potential. However, this small dependence on the forward propagation distance does not change our conclusions regarding the convergence of the results towards infinite regions.

\begin{figure}[t]
  \begin{center}
  \includegraphics[scale=0.45]{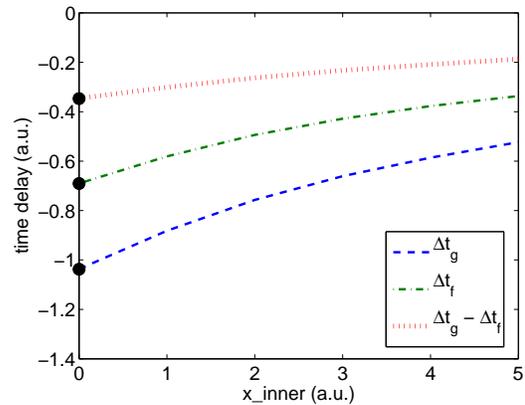}
  \end{center}
  \caption{\small (Color online) Time delays and difference between time delays as a function of inner integration boundary
$x_{\text{inner}}$. Symbols and laser parameters are the same as in Fig. \ref{delay_outer_boundary}. We also plotted the WS time delays as black dots in this figure.}
\label{delay_inner_boundary}
\end{figure}

We also note from the results in Fig.\ \ref{delay_outer_boundary} that the time delay increases most strongly in the
region close to the center of the potential, where the potential changes
%dramatically
most strongly. This indicates that the results
should depend on the choice of the inner boundary $x_\text{inner}$ of the region $R$. To study this feature,
we fixed the outer boundary of $R$ at $x_\text{outer}=500$, which is large enough to obtain converged results in case
of the Yukawa potential, and then
varied the inner boundary $x_\text{inner}$. The results in Fig. \ref{delay_inner_boundary} show the expected
dependence on the choice of $x_\text{inner}$, the absolute values of the time delays decrease by half as $x_\text{inner}$
increases from $0 \; \text{a.u.}$ to $5\;\text{a.u.}$.
In the remainder of the present studies we have chosen the $x_\text{inner}=0\;\text{a.u.}$ as the inner boundary, since
this value corresponds to the expectation value of $x$ for all the bound states investigated here.

\subsection{Wigner-Smith time delay for photoionization in a short-range potential}

In the case of the Yukawa potential, we also compared our numerical results, obtained in the time-dependent numerical simulations, with calculations
of the WS time delay as a derivative of the induced phase shift, c.f.\ Eq.\ (\ref{WS_delay}), obtained from a time-independent
scattering approach.
In order to obtain the latter for photoionization by a light pulse with finite duration, we first considered
the scattering of an electron, incident from $x = -\infty$ with a momentum $k$, of the short-range Yukawa potential.
We solved the corresponding time-independent Schr\"odinger equation numerically using the fourth order Runge-Kutta method up to $|x| = 500$, projected the
numerical solution onto the appropriate plane-wave solutions for $x \rightarrow \pm\infty$, and obtained the WS time delay
for the scattering process $\Delta t_{\text{WS}}^{\text{(scat)}}$ as the derivative of
the phase shift in the plane wave propagating in positive $x$-direction with respect to the energy of the incident particle.
In order to take account of the energy spread of the ionizing wave packet in a specific photoionization process, we averaged
$\Delta t_{\text{WS}}^{\text{(scat)}}$ over the energy spectrum of the wave packet, as obtained in our time-dependent numerical simulations.
Finally, we considered the photoionization as a half-scattering process and divided the result of the average by two.
The resulting WS time delays for photoionization are shown as black dots in Fig.\ \ref{delay_inner_boundary} and are
in good agreement with our numerical results, obtained from the time-dependent calculations, for $x_\text{inner} = 0$ and $x_\text{outer}=500$.
This is in support of the applicability of our approach to obtain time delays from the time-dependent numerical simulations.

\subsection{Dependence of time delay on XUV pulse parameters}

\begin{figure}[t]
  \begin{center}
  \includegraphics[scale=0.45]{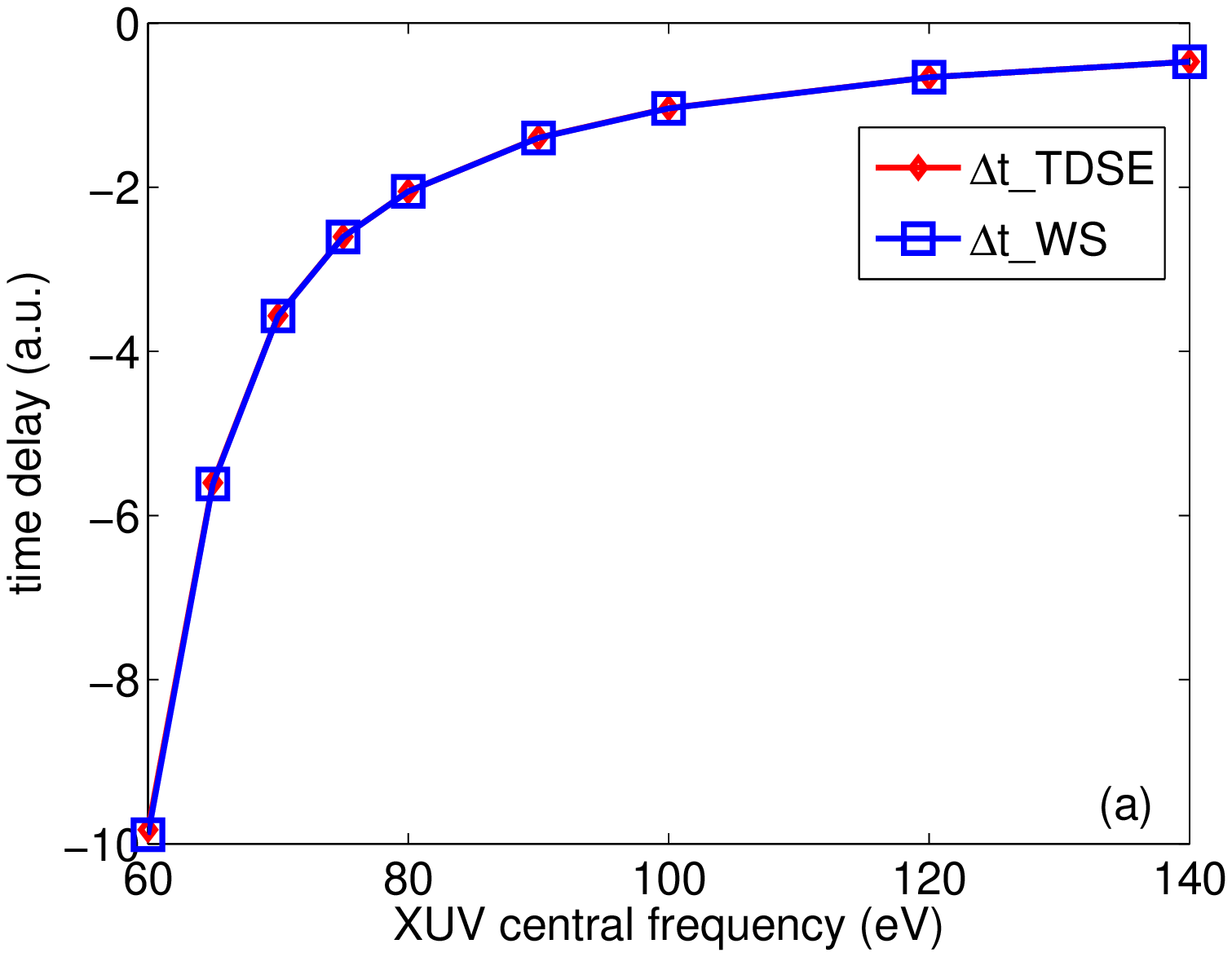}
  \includegraphics[scale=0.45]{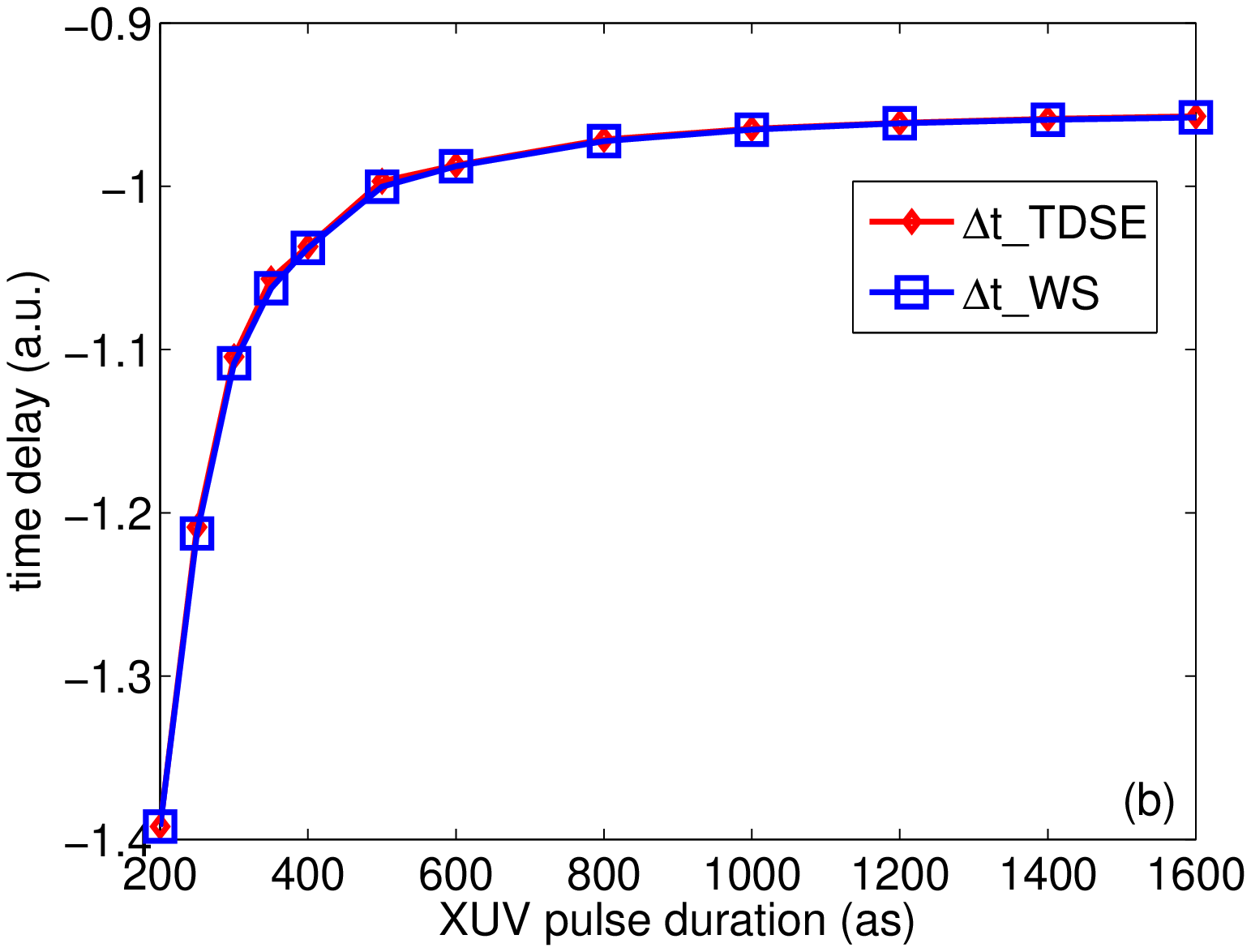}
  \end{center}
  \caption{\small
(Color online) Time delays for ionization from the ground state of the Yukawa potential as functions of (a) the XUV photon frequency ($\tau = 400$ as) and
(b) the pulse duration of the XUV pulse ($\omega = 100$ eV). Comparison between results from the present
TDSE calculations (red diamonds) and those for the Wigner Smith time delay (blue open squares) is shown.
Other laser parameters were: $I=1\times10^{15}$ W/cm$^2$ and $\phi=0$.
}
\label{delay_omega_pulseduration}
\end{figure}

Next, we studied the dependence of the time delay introduced here on the parameters of the XUV ionizing pulse for photoionization
from the ground state of the Yukawa potential.
In Fig.\ \ref{delay_omega_pulseduration} we present our results as functions of (a) the XUV frequency at a fixed pulse duration of $\tau=400$ as
and (b) the duration of the XUV pulse at a fixed frequency of $\omega = 100$ eV. The peak intensity was $I=1\times 10^{15}$ W/cm$^2$ and
the CEP was $\phi =0$ in each of these simulations.

The results agree well with qualitative expectations. The absolute value of the time delay decreases towards zero as the frequency of the ionizing
XUV pulse and, hence, the final kinetic energy of the emitted electron increases (cf.\ panel a). This is due to the fact that the effect
of the potential on the motion of the electron becomes negligible in the limit of infinitely large kinetic energy of the electron (i.e.\ infinite
large XUV frequency) and therefore the time spent in the potential approaches that of the free particle in this limit.

We further find that the
absolute value of the time delay decreases with an increase of the XUV pulse duration (panel b). This dependence is closely related to that
presented in Fig.\ \ref{delay_omega_pulseduration}(a) and can be qualitatively understood as follows. Due to the finite pulse duration the ionized electron wave packet has a certain
bandwidth about a central kinetic energy. Consequently, the time delay obtained for the wave packet can be considered as an average over contributions
at particular electron energies within the bandwidth (weighted by the ionization probability at a given energy). As indicated by the results in panel (a)
the time delay does not change linearly with the kinetic energy. Therefore, the time delay obtained for a wave packet will be smaller than its contribution at the central kinetic energy
or the expectation value of the kinetic energy. This difference decreases and, thus, the time delay for the wave packet increases as the energy bandwidth of the wave packet decreases,
i.e.\  as the pulse duration increases.
Furthermore, it is found that the expectation value of the kinetic energy of the ionizing wavepacket increases as the XUV pulse duration increases,
which also causes the time delay to increase.

For each of the results from the time-dependent simulations presented in Fig.\ \ref{delay_omega_pulseduration} we also
calculated the WS time delay for photoionization, as described in the previous subsection. Again, we found excellent agreement between
the results from the time-independent scattering approach (blue open squares in Fig.\ \ref{delay_omega_pulseduration}) and our numerical time-dependent simulations.
We may note parenthetically, that our simulations results also agree well with those
of classical calculations (not shown) in which the time delays are determined by
\begin{equation}
\Delta t_\text{classical} =
\int_{0}^{x_\text{outer}} \frac{1}{\sqrt{2(\langle E\rangle-V(x))}} dx-\frac{x_\text{outer}}{\sqrt{2\langle E \rangle}}.
\label{delay_classical}
\end{equation}
where $\langle E \rangle$ is the expectation value of the kinetic energy $E$ of the electron for a given ionized wave packet.

\section{Streaking of photoionization processes by near-infrared field}

In an attosecond streaking experiment \cite{itatani02},
an IR field is used to map time information to the momentum space. One of the questions that arise in this context is whether or not
the streaking field influences the observed quantities. Although the time delays introduced here do not necessarily correspond to those observed in recent streaking experiments, we can show, in general, how the effect of the streaking field can be studied with our present method. To this end, we will first discuss how
the streaking field can be included in the numerical simulations of time delays and then study the impact of a streaking IR field on the time delays for the
short-range Yukawa as well as the long-range Coulomb potential by varying the parameters of the streaking field.

\subsection{Streaking field in the numerical simulation}

The streaking field is represented by an additional potential $V_\text{streak}$ which we consider as part of the potential $V({\bm r})$ in Eq.\ (\ref{schroedinger}).
Thus, in the present calculations the streaking field is considered on equal footing with the
atomic potential. After (forward) propagation of the wavefunction from its initial state and separation of the
bound and ionizing part of the wavefunction, we then propagate the ionizing part of the wavefunction backwards once within the combination of the
atomic potential and the streaking field and once as a free particle.

As a result we obtain the time delay, associated with the ionizing part of the wavefunction in the combined potential of the short- or long-range interaction
and the streaking field as
\begin{equation}
\Delta t^{(\text{IR})}_{\Psi_i,R} = t_{\Psi_i,R}^\text{(IR)} - t_{\Psi^{(0)},R}
\label{num_delay_IR}
\end{equation}
where $t_{\Psi_i,R}^\text{(IR)}$ is the time the ionizing wave packet spends in region $R$ in the presence of the (short- or long-range) potential and the
IR streaking field, and $t_{\Psi^{(0)},R}$ is the time for the free particle case.

Here, the forward propagation of the wavefunction has to be continued as long as the IR streaking field is present.
As noted above, for the long-range Coulomb potential the time delay depends on the distance the wavepacket is propagated in forward direction. In order to
keep the corresponding error small in our current analysis we used a large grid of $-13000$ a.u. to $13000$ a.u. and
stopped the forward propagation when the expectation value of the ionizing wave packet reaches 8000 a.u.. We increased the spatial step to $\delta x=0.1$
and the time step to $\delta t=0.02$ as compared to the previous calculations. Test calculations showed that the relative error of the present results is about 1\%.
In order to compare with the results, presented above, we considered the same parameters for the Yukawa and the Coulomb potential as before. We chose the
first excited state as the initial state and checked that the ionization induced by the IR field is negligible up to a streaking field intensity of $1\times10^{13}$ W/cm$^2$,
which is large enough for the streaking purpose.

\subsection{Effects of the probing pulse on time delays}

\begin{figure}[t]
  \begin{center}
  \includegraphics[scale=0.255]{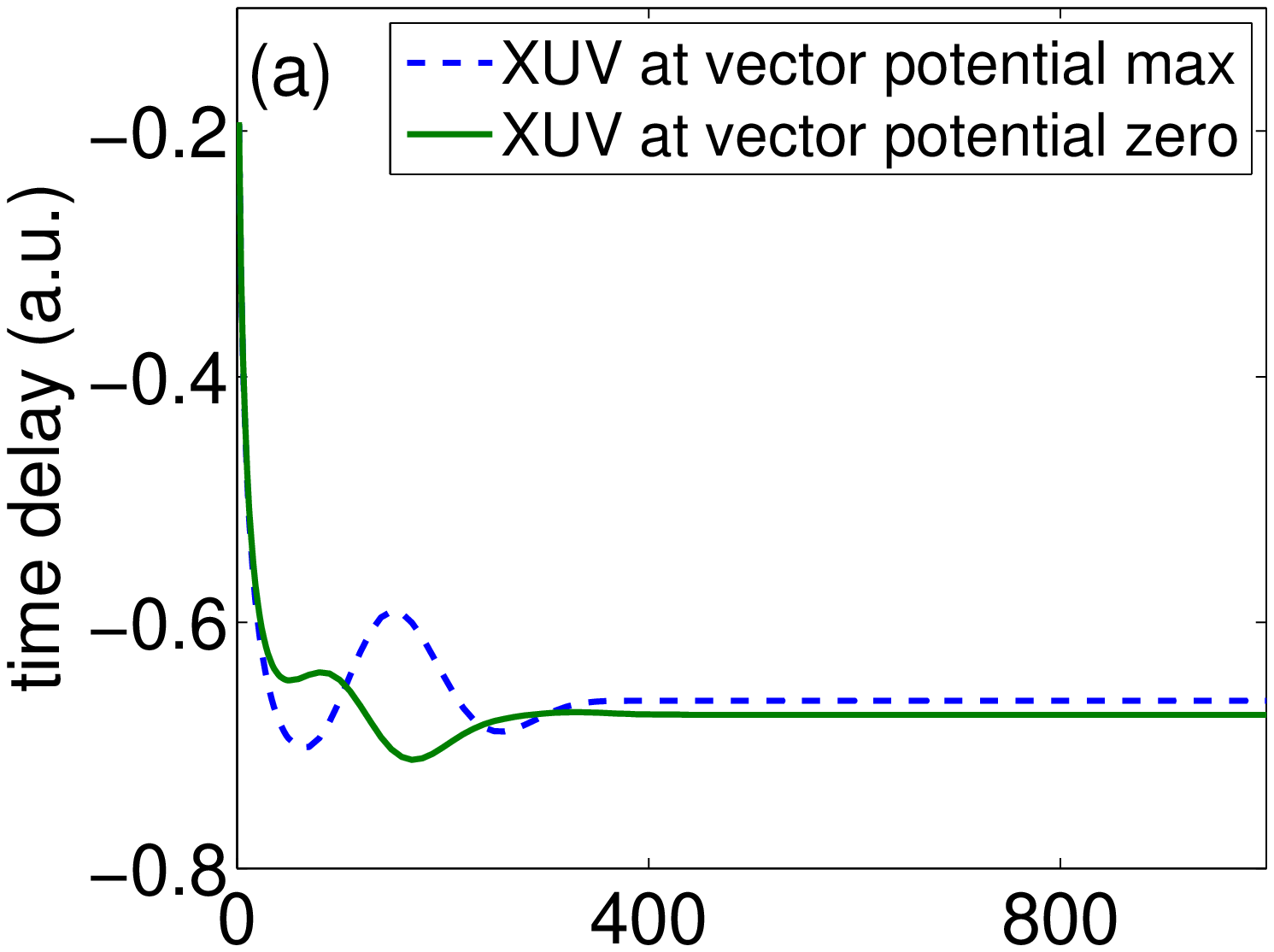}
  \includegraphics[scale=0.255]{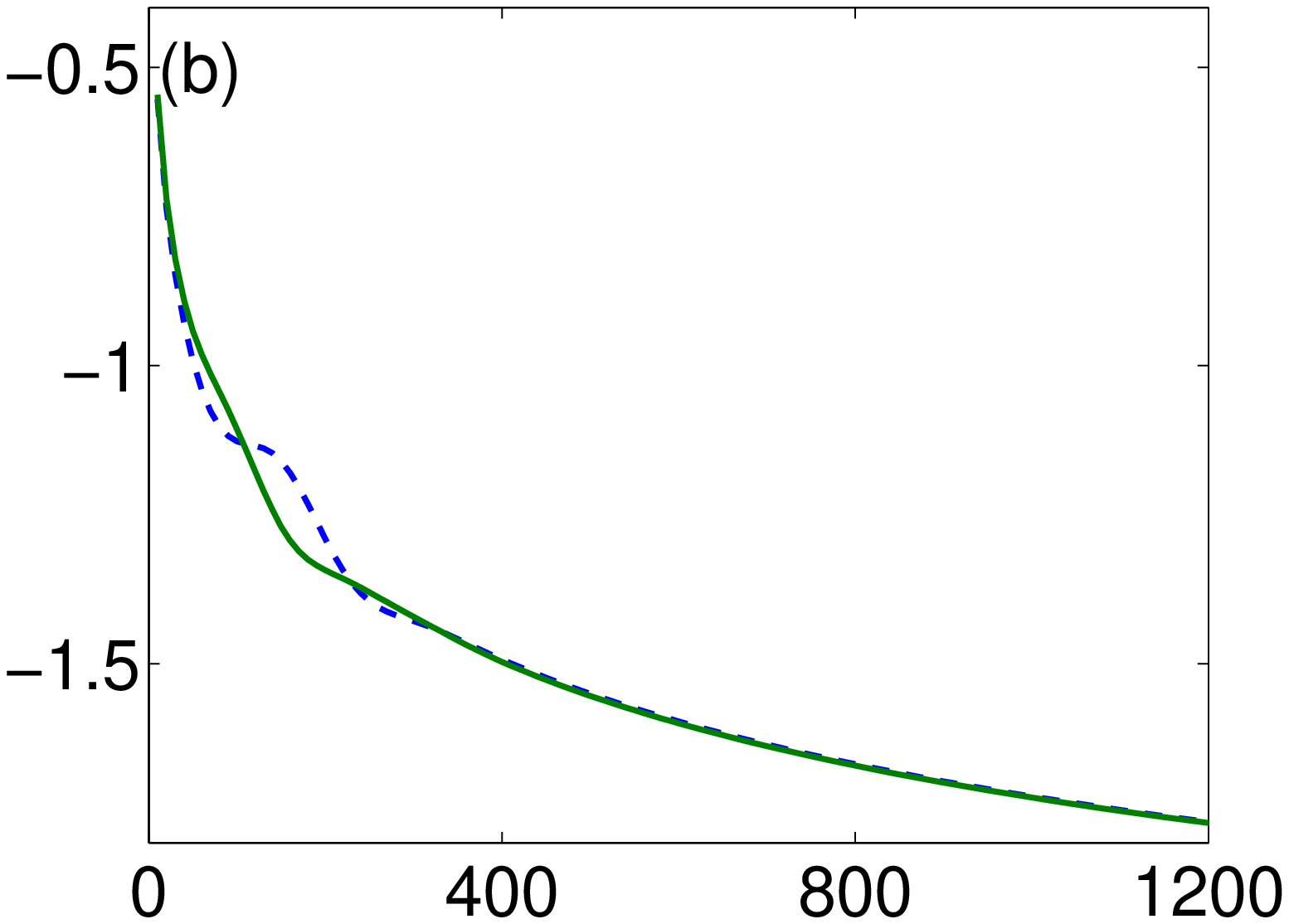}\\
  \includegraphics[scale=0.255]{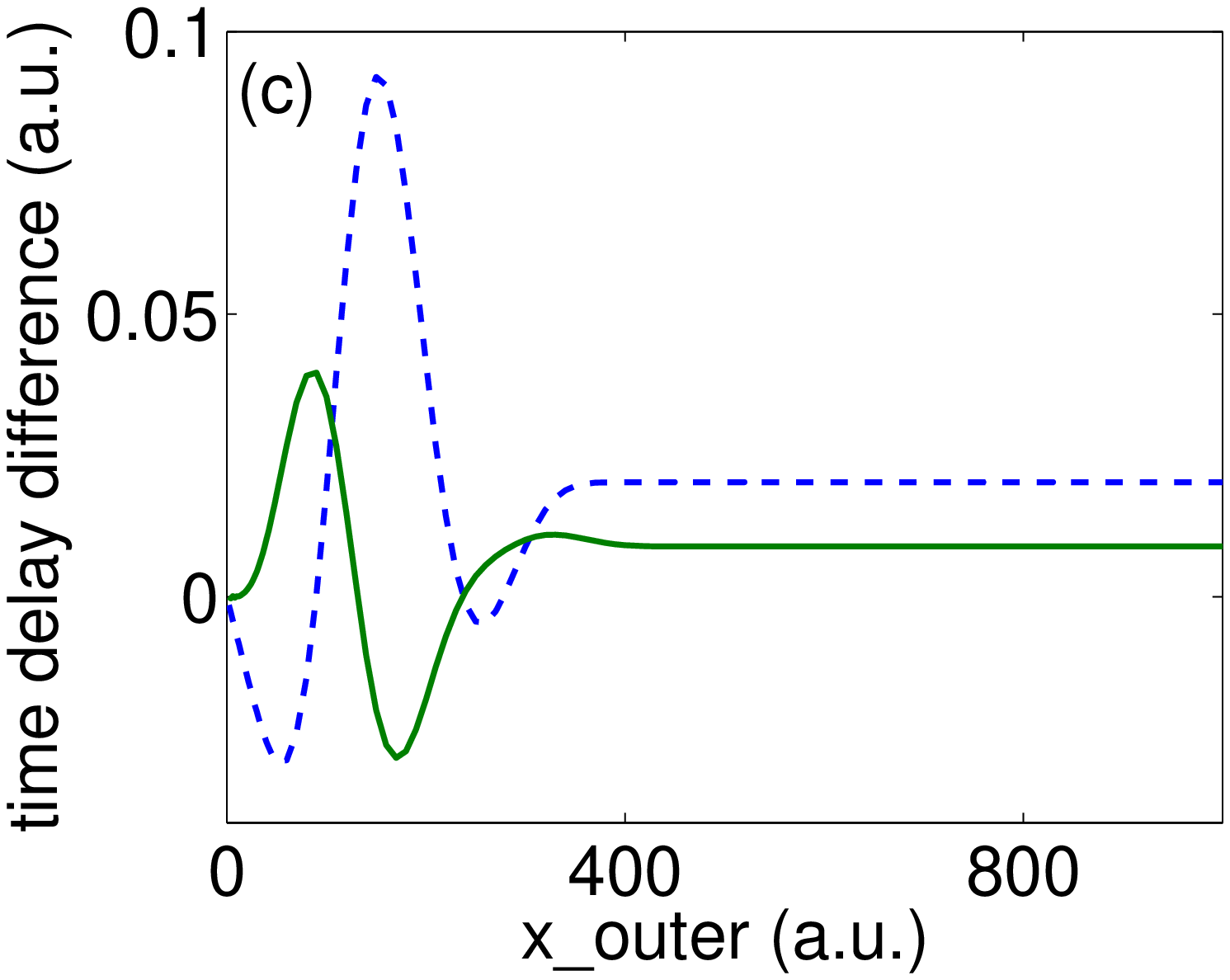}
  \includegraphics[scale=0.255]{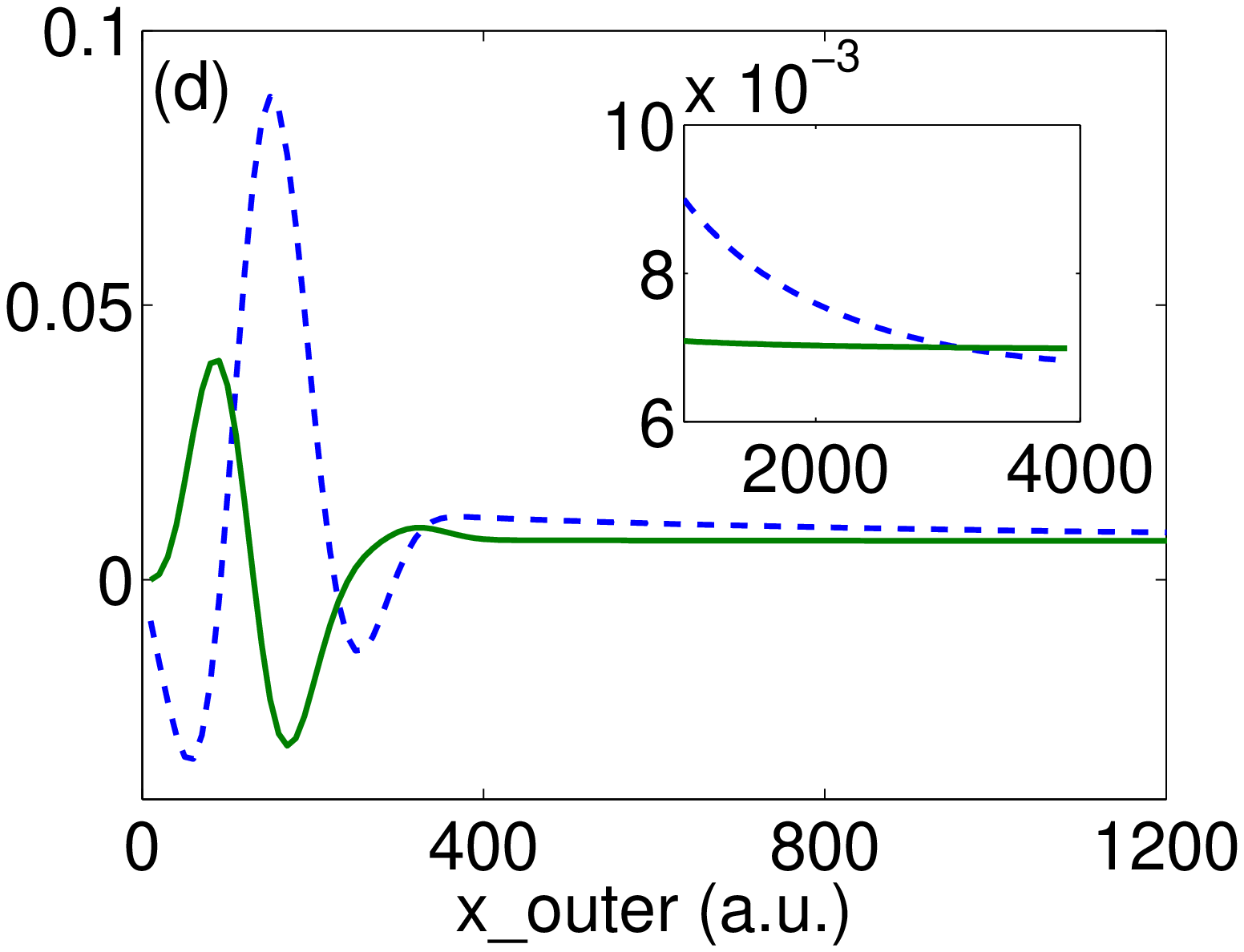}
  \end{center}
  \caption{\small (Color online) Time delays (upper row) and time delay differences (lower row) as a function of outer boundary $x_\text{outer}$ of $R$ for Yukawa potential (left column) and Coulomb potential (right column). For each potential we have centered the XUV pulse at two different positions which correspond to the maximum (blue dash-dotted line) and zero (green solid line) of the IR vector potential respectively. The XUV parameters are: $I_\text{XUV}=1\times10^{15}$ W/cm$^2$, $\omega_\text{XUV}=100$ eV, $\tau_\text{XUV}=400$ as, and $\phi_\text{XUV}=0$. The IR parameters are: $I_\text{IR}=1\times10^{12}$ W/cm$^2$, $\lambda_\text{IR}=800$ nm, $N_\text{IR}=3$ cycle, and $\phi_\text{IR}=0$. The small box in (d) shows the long-range behavior of the two curves.}
\label{IR_delay_R}
\end{figure}

In order to analyze the effect of the IR probing field we obtained the time delay for photoionization from the first excited state of the
Yukawa as well as the Coulomb potential in the streaking field. To this end, we applied the ionizing XUV pulse centered at the maximum of
the IR streaking field (zero of the vector potential) as well as centered at the central zero of the IR streaking field (maximum of the
vector potential). In the upper row of Fig.\ \ref{IR_delay_R} the results for the
time delays in the streaking field (green solid lines: XUV centered at zero of
the vector potential; blue dashed lines: XUV centered at maximum of vector potential) are shown as a function of the outer boundary $x_\text{outer}$
of the region $R$ for the Yukawa (left) and the Coulomb potential (right).
As in the results without streaking field, we see that there is a well-defined limit of the time delays for the short-ranged Yukawa potential as
the region $R$ increases, while there is no convergence found for the Coulomb potential.

In order to see the effect of the probing field, we present in the lower row of Fig.\ \ref{IR_delay_R}
the difference between the time delays in the streaking field to those without streaking field as a function of the outer boundary of the region $R$, i.e.\
\begin{equation}
\Delta T = \Delta t^{(\text{IR})}_{{\Psi_i},R} - \Delta t_{{\Psi}_i,R},
\label{num_delay_difference}
\end{equation}
Although there is no well-defined limit of the time delays for infinite regions in the Coulomb case, neither with nor without streaking field,
for any finite region the time delays introduced here are well-defined and the effect of the streaking field can be analyzed. The same argument applies to the
weak dependence of the Coulomb results on the distance of forward propagation in our simulations.

For both potentials, we see that the time delay difference $\Delta T$ oscillates for $x_\text{outer} < 400$.
This oscillation is due to the presence of the IR field, since the ionized wave packet propagated up to about $x \simeq 400$ before the
IR streaking field ceased in the present simulations. We note that the differences $\Delta T$ are small, less than 3\% for the Yukawa potential
and less than the numerical error of 1\% for the Coulomb potential, compared to the time delays induced by the atomic potentials themselves.
In the present calculations we therefore do not find a significant effect of the streaking field, neither for a short-range nor for
a long-range potential.
%This is in contrast with the findings of some previous theoretical work \cite{ivanov11,zhang10,nagele12}\textcolor{blue}{, which is under our expectation since our introduced time delays, as pointed out before, are not necessarily equivalent to the time delays measured in the streaking experiments.}

Before we continue to further study the influence of the IR streaking field, we note a subtle point in the results obtained for
the Coulomb potential, which are presented in Fig.\ \ref{IR_delay_R}. While neither the time delays with and without streaking field converge
as a function of the outer boundary $x_\text{outer}$, we find a converged result (within the numerical error) for the time delay difference
$\Delta T$,
if the XUV pulse is centered about the zero of the vector potential of the streaking field (see green solid line in Fig.\ \ref{IR_delay_R}(d)).
This occurs since in this case
the momentum distribution of the ionizing wave packet at the end of the forward propagation is the same as that of the no streaking field case.
In contrast, if the XUV pulse is applied at the maximum of the vector potential of the streaking field, the final momentum distribution is
shifted and thus no convergence of the time delay difference within the range of present boundaries is found
(see blue dashed line in the inset of Fig.\ \ref{IR_delay_R}(d)).

\begin{figure}[t]
  \begin{center}
  \includegraphics[scale=0.35]{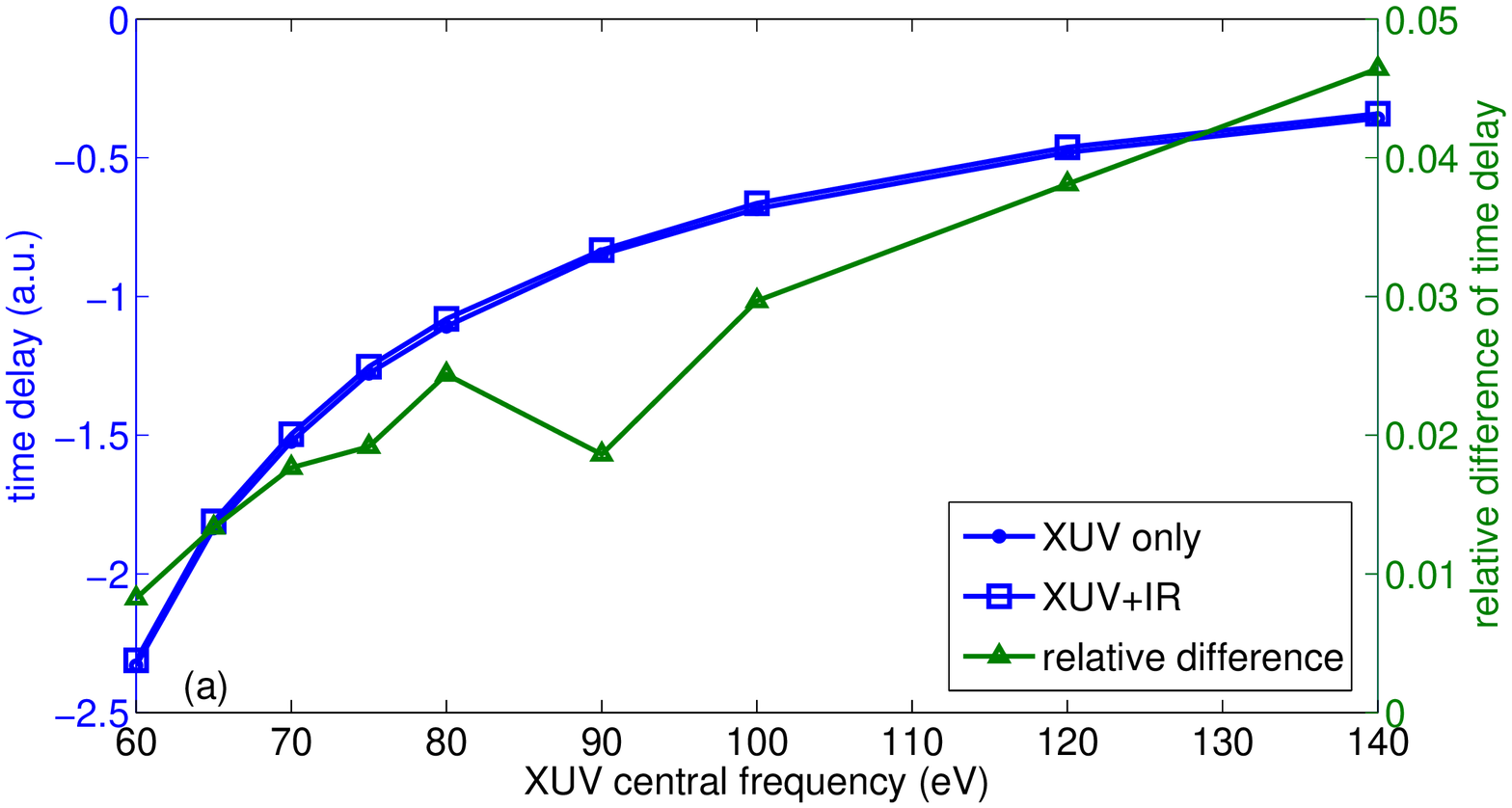}
  \includegraphics[scale=0.35]{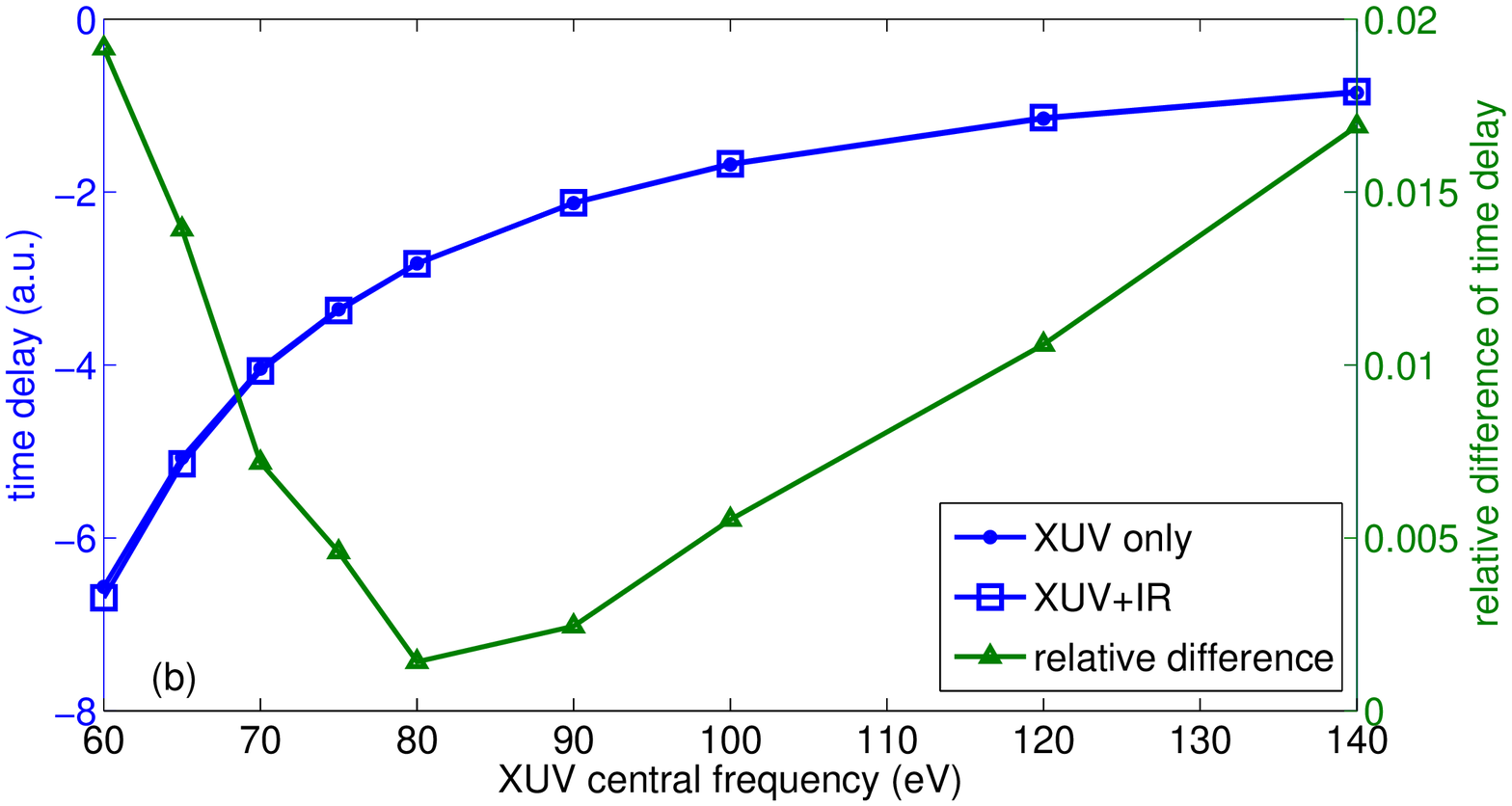}
  \end{center}
  \caption{\small
(Color online) Time delays with (blue open squares) and without (blue filled circles) IR streaking field as well as
relative differences between the results (green filled triangles) as a function of XUV central frequency for
(a) Yukawa potential and (b) Coulomb potential. The XUV pulse was centered at the middle (maximum vector potential) of the IR field.
Other laser parameters are the same as in Fig.\ \ref{IR_delay_R}. For the Coulomb case the time delays are calculated at $x_\text{outer}=800$.}
	\label{IR_delay_XUV_omega}
\end{figure}

\begin{figure}[t]
  \begin{center}
  \includegraphics[scale=0.45]{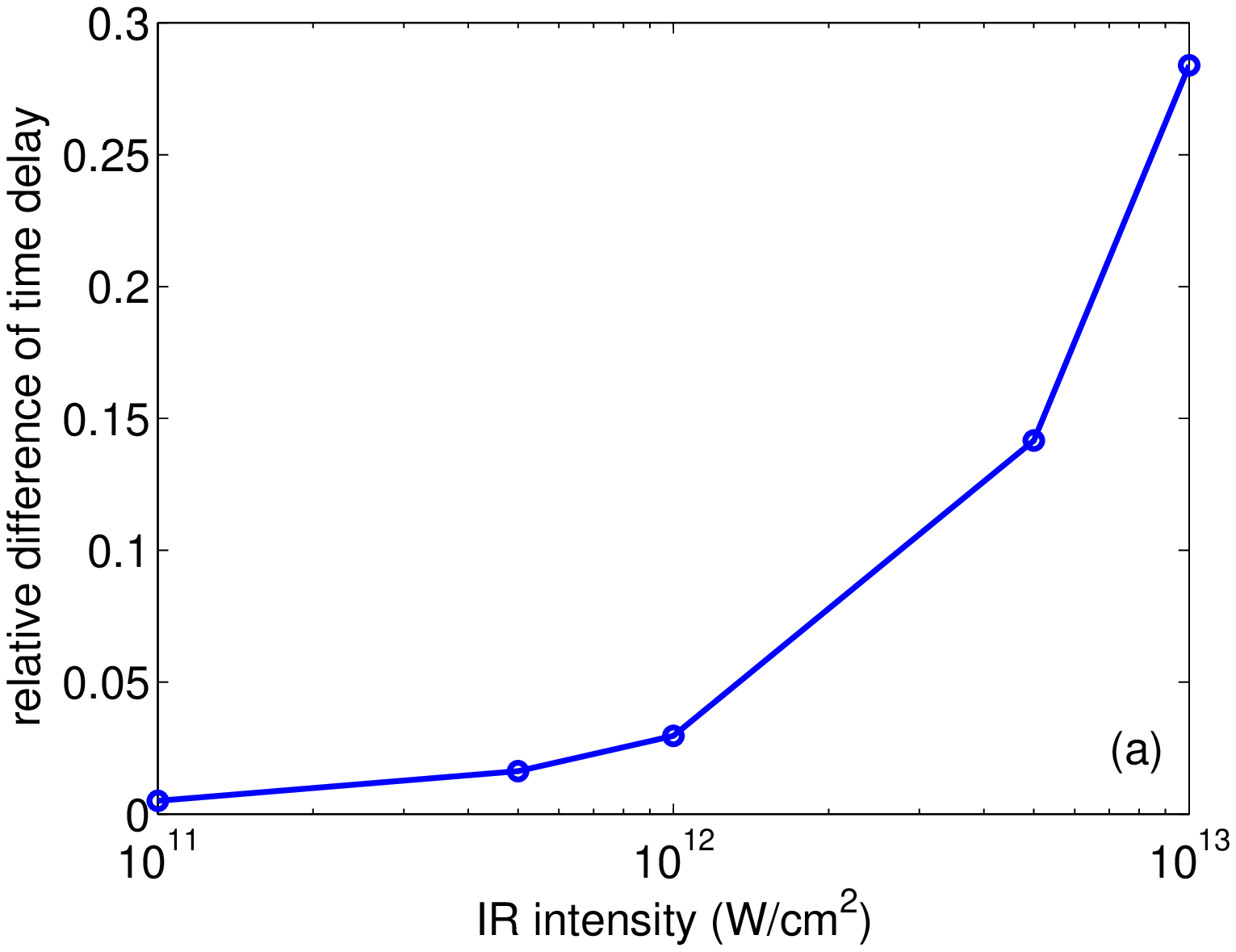}
  \includegraphics[scale=0.45]{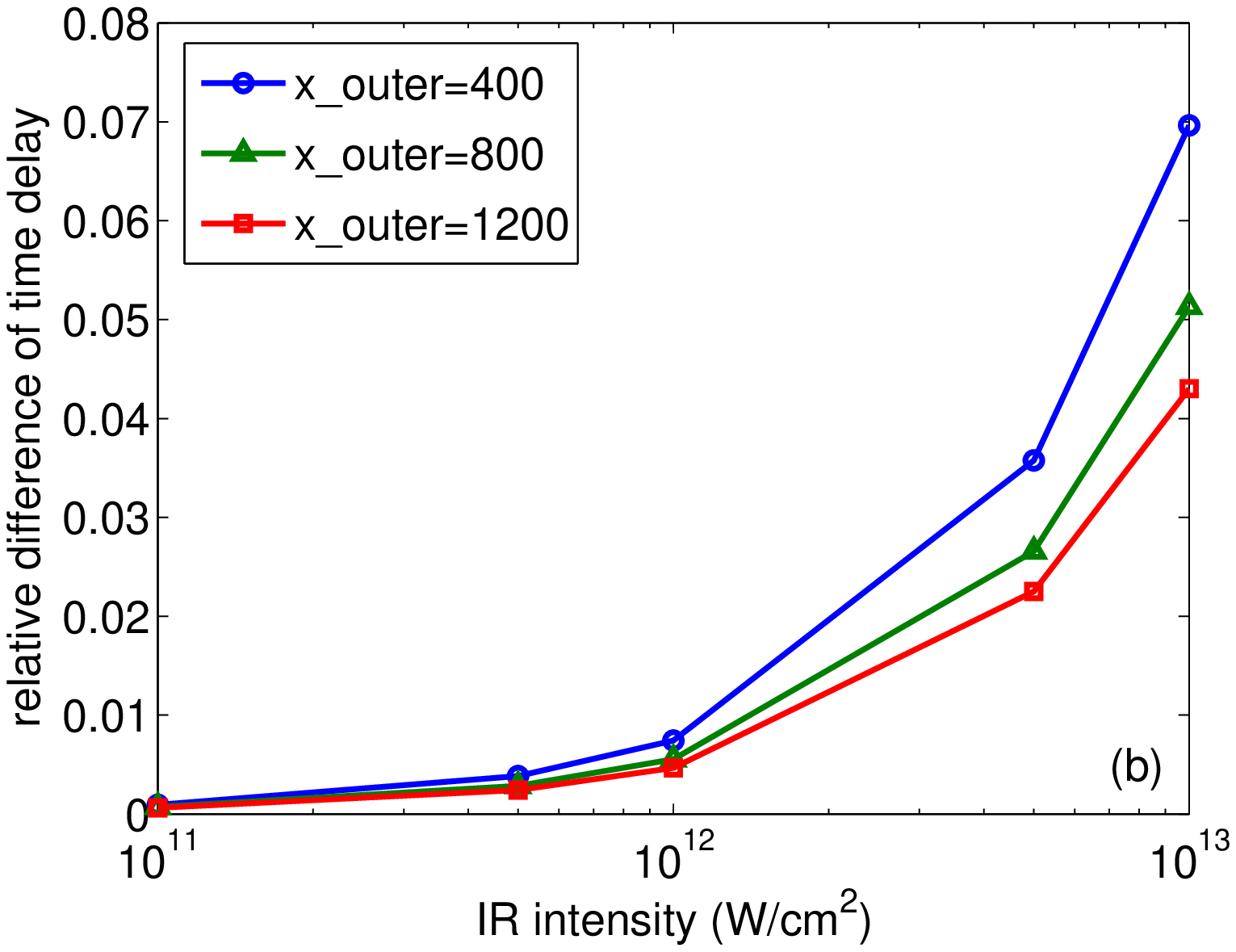}
  \end{center}
  \caption{\small (Color online) Relative differences of time delay as function of IR intensity for (a) Yukawa potential and (b) Coulomb potential. The XUV pulse is centered at the middle of the IR field. Laser parameters are the same as in Fig.\ \ref{IR_delay_R} expect $I_\text{IR}$ is changing. For the Coulomb case the time delays are taken at $x_\text{outer}=400$, $800$ and $1200$.}
	\label{delay_difference_IR_intensity}
\end{figure}

The conclusion that the streaking field does not influence the time delay introduced here significantly holds over a large range of XUV frequencies as well as
for intensities of the streaking field up to about $10^{13}$ W/cm$^2$. In Fig.\ \ref{IR_delay_XUV_omega} we present the results for the time delay
obtained in (a) the Yukawa potential and (b) the Coulomb potential with (blue open squares) and without (blue filled circles) streaking field
as a function of the XUV frequency. Since the results are in close agreement we also show the relative difference between them (green filled triangles),
which does not exceed 5\% and 2\% in the Yukawa and Coulomb case, respectively.

As one would expect, the relative difference between the results for the time delay obtained with and without streaking field do increase
with an increase of the IR streaking field intensity. This can be clearly seen from the results shown in  Fig.\ \ref{delay_difference_IR_intensity}.
It appears that for IR intensities up to $10^{12}$ W/cm$^2$ the relative difference between the results is small enough such that there is
no significant effect on the time delay. While the relative difference quickly increases beyond 10\% in the case of the Yukawa potential with
a further increase of the IR intensity, the 10\%-limit is not reached for an IR intensity of $10^{13}$ W/cm$^2$ in the case of the Coulomb potential.

\section{Conclusion}

In summary, we have applied a fundamental definition of time delay to time-dependent numerical simulations on the grid. To this end, we have obtained the difference
between the time a particle spends in a finite region of a potential and the time a free particle spends in the same region using a back-propagation technique.
Our method expands the options for a theoretical analysis of ultrashort time-dependent processes. For any finite region in space the time delay introduced here is well-defined, even for long-range potentials, and time delays can be determined as a function of time after the emission of the photoelectron.
The method is applied to photoionization of an electron in a short-range Yukawa as well as a long-range Coulomb potential by an attosecond
XUV pulse. It is found that the numerical results are in excellent agreement with those for the (asymptotic) WS time delay, obtained
as the derivative of the phase shift with respect to the energy, for the short-range potential. In contrast, the numerical results in the case of the Coulomb
potential are finite for any finite region, but they do not converge as the region increases to infinity, as expected. The well-defined time delays for
a finite region enabled us to study the impact of a near-infrared streaking (or probing) pulse for both potentials.
For the time delays introduced in this paper, our results show that the effect is small as long as the intensity of the probing field is below $10^{13}$ W/cm$^2$.

\begin{acknowledgments}
J.S. and A.B. acknowledge financial support by the U.S. Department of Energy,
Division of Chemical Sciences, Atomic, Molecular and Optical Sciences Program.
H.N. was supported via a grant from the U.S. National Science Foundation (Award No. PHY-0854918).
A.J.-B. acknowledges financial support by the U.S. National Science Fundation (Award No. PHY-1068706).
This work utilized the Janus supercomputer, which is supported by the U.S.\ National Science Foundation (award number CNS-0821794) and the University of Colorado Boulder. The Janus supercomputer is a joint effort of the University of Colorado Boulder, the University of Colorado Denver and the National Center for Atmospheric Research.
\end{acknowledgments}


\begin{thebibliography}{99}
%
\bibitem{drescher02}
M. Drescher, M. Hentschel, R. Kienberger, M. Uiberacker, V. Yakovlev, A. Scrinzi, T. Westerwalbesloh, U. Kleineberg, U. Heinzmann, and F. Krausz,
Nature (London) {\bf 419}, 803 (2002).
%

%
\bibitem{goulielmakis10}
E. Goulielmakis, Z. -H. Loh, A. Wirth, R. Santra, N. Rohringer, V. S. Yakovlev, S. Zherebtsov, T. Pfeifer, A. M. Azzeer, M. F. Kling, S. R. Leone, and F. Krausz,
Nature (London) {\bf 466}, 739 (2010).
%

%
\bibitem{uiberacker07}
M. Uiberacker, T. Uphues, M. Schultze, A. J. Verhoef, V. Yakovlev, M. F. Kling, J. Rauschenberger, N. M. Kabachnik, H. Schrder, M. Lezius, K. L. Kompa, H. G. Muller, M. J. J. Vrakking, S. Hendel, U. Kleineberg, U. Heinzmann, M. Drescher, and F. Krausz,
Nature (London) {\bf 446}, 627 (2007).
%

%
\bibitem{gagnon07}
E. Gagnon, P. Ranitovic, X.-M. Tong, C. L. Cocke, M. M. Murnane, H. C. Kapteyn, and A. S. Sandhu,
Science {\bf 317}, 1374 (2007).
%

%
\bibitem{cavalieri07}
A. L. Cavalieri, N. M\"{u}ller, T. Uphues, V. S. Yakovlev, A. Baltu\v{s}ka, B. Horvath, B. Schmidt, L. Bl\"{u}mel, R. Holzwarth, S. Hendel, M. Drescher, U. Kleineberg, P. M. Echenique, R. Kienberger, F. Krausz, and U. Heinzmann,
Science {\bf 449}, 1029 (2007).
%

%
\bibitem{schultze10}
M. Schultze, M. Fie\ss, N. Karpowicz, J. Gagnon, M. Korbman, M. Hofstetter, S. Neppl, A. L. Cavalieri, Y. Komninos, T. Mercouris, C. A. Nicolaides, R. Pazourek, S. Nagele, J. Feist, J. Burgd\"{o}rfer, A. M. Azzeer, R. Ernstorfer, R. Kienberger, U. Kleineberg, E. Goulielmakis, F. Krausz, and V. S. Yakovlev,
Science {\bf 328}, 1658 (2010).
%

%
\bibitem{klunder11}
K. Kl\"{u}nder, J. M. Dahlstr\"{o}m, M. Gisselbrecht, T. Fordell, M. Swoboda, D. Gu\'{e}not, P. Johnsson, J. Caillat, J. Mauritsson, A. Maquet, R. Ta\"{i}eb, and A. L'Huillier,
Phys. Rev. Lett. {\bf 106}, 143002 (2011).
%

%
\bibitem{kheifets10}
A. S. Kheifets and I. A. Ivanov,
Phys. Rev. Lett. {\bf 105}, 233002 (2010).
%

%
\bibitem{zhang10}
C.-H. Zhang and U. Thumm,
Phys. Rev. A {\bf 82}, 043405 (2010).
%

%
\bibitem{nagele11}
S. Nagele, R. Pazourek, J. Feist, K, Doblhoff-Dier, C. Lemell, K. T\"{o}k\'{e}si, and J. Burgd\"{o}rfer,
J. Phys. B {\bf 44}, 081001 (2011).
%

%
\bibitem{ivanov11}
M. Ivanov and O. Smirnova,
Phys. Rev. Lett. {\bf 107}, 213605 (2011).
%

%
\bibitem{moore11}
L. R. Moore, M. A. Lysaght, J. S. Parker, H. W. van der Hart, and K. T. Taylor,
Phys. Rev. A {\bf 84}, 061404 (2011).
%

%
\bibitem{zhang11}
C.-H. Zhang and U. Thumm,
Phys. Rev. A {\bf 84}, 033401 (2011).
%

%
\bibitem{kheifets11}
A. S. Kheifets, I. A. Ivanov, and I. Bray,
J. Phys. B {\bf 44}, 101003 (2011).
%

%
\bibitem{ivanov11pra}
I. A. Ivanov,
Phys. Rev. A {\bf 83}, 023421 (2011).
%

%
\bibitem{ivanov12}
I. A. Ivanov,
Phys. Rev. A {\bf 86}, 023419 (2012).
%

%
\bibitem{sukiasyan12}
S. Sukiasyan, K. L. Ishikawa, and M. Ivanov,
Phys. Rev. A {\bf 86}, 033423 (2012).
%

%
\bibitem{dahlstrom12}
J. M. Dahlstr\"{o}m, A. L'Huillier, and A. Maquet,
J. Phys. B {\bf 45}, 183001 (2012).
%

%
\bibitem{spiewanowski12}
M. D. Spiewanowski and L. B. Madsen,
Phys. Rev. A {\bf 86}, 045401 (2012).
%

%
\bibitem{guenot12}
D. Gu\'{e}not, K. Kl\"{u}nder, C. L. Arnold, D. Kroon, J. M. Dahlstr\"{o}m, M. Miranda, T. Fordell, M. Gisselbrecht, P. Johnsson, J. Mauritsson, E. Lindroth, A. Maquet, R. Ta\"{i}eb, A. L'Huillier, and A. S. Kheifets,
Phys. Rev. A {\bf 85}, 053424 (2012).
%

%
\bibitem{pazourek12}
R. Pazourek, J. Feist, S. Nagele, and J. Burgd\"{o}rfer,
Phys. Rev. Lett. {\bf 108}, 163001 (2012).
%

%
\bibitem{nagele12}
S. Nagele, R. Pazourek, J. Feist, and J. Burgd\"{o}rfer,
Phys. Rev. A {\bf 85}, 033401 (2012).
%

%
\bibitem{wigner55}
E. P. Wigner,
Phys. Rev. {\bf 98}, 145 (1955).
%

%
\bibitem{smith60}
F. T. Smith,
Phys. Rev. {\bf 118}, 349 (1960).
%

%
\bibitem{newton_book}
R. G. Newton,
\textit{Scattering Theory of Waves and Particles} (Springer-Verlag, New York, 1982), chap.11.
%

%
\bibitem{ho83}
Y. K. Ho,
Phys. Rep. {\bf 99}, 1 (1983).
%

%
\bibitem{mccurdy04}
C. W. McCurdy, M. Baertschy, and T. N. Rescigno,
J. Phys. B {\bf 37}, R137 (2004).
%

%
\bibitem{itatani02}
J. Itatani, F. Qu\'{e}r\'{e}, G. L. Yudin, M. Y. Ivanov, F. Krausz, and P. B. Corkum,
Phys. Rev. Lett. {\bf 88}, 173903 (2002).
%

\end{thebibliography}
\end{document}